\documentclass[11pt]{article}
\usepackage{epsfig}
\usepackage{latexsym}
\usepackage{graphicx}
\usepackage{amsmath,amssymb}

\def\ltsim{\lower3pt\hbox{$\, \buildrel \frac{<}{\sim} \, $}}  
\def\gtsim{\lower3pt\hbox{$\, \buildrel \frac{>}{\sim} \, $}}  
\makeatletter
\def\section{\@startsection {section}{1}{\z@}{-3.5ex plus -1ex minus
 -.2ex}{2.3ex plus .2ex}{\large\bf}}
\def\subsection{\@startsection{subsection}{2}{\z@}{-3.25ex plus -1ex
minus -.2ex}{1.5ex plus .2ex}{\normalsize\bf}}
\makeatother
\newcommand{\captionfonts}{\small}
\makeatletter  % Allow the use of @ in command names
\long\def\@makecaption#1#2{%
  \vskip\abovecaptionskip
  \sbox\@tempboxa{{\captionfonts #1: #2}}%
  \ifdim \wd\@tempboxa >\hsize
    {\captionfonts #1: #2\par}
  \else
    \hbox to\hsize{\hfil\box\@tempboxa\hfil}%
  \fi
  \vskip\belowcaptionskip}
\makeatother   % Cancel the effect of \makeatletter
\catcode`@=11
\def\marginnote#1{}
\newcount\hour
\newcount\minute
\newtoks\amorpm
\hour=\time\divide\hour
by60
\minute=\time{\multiply\hour by60 \global\advance\minute
by-\hour}
\edef\standardtime{{\ifnum\hour<12 \global\amorpm={am}
\else\global\amorpm={pm}\advance\hour by-12 \fi
 \ifnum\hour=0
\hour=12 \fi
 \number\hour:\ifnum\minute<10
0\fi\number\minute\the\amorpm}}
\edef\militarytime{\number\hour:\ifnum\minute<10
0\fi\number\minute}
\def\draftlabel#1{{\@bsphack\if@filesw
{\let\thepage\relax
 \xdef\@gtempa{\write\@auxout{\string
\newlabel{#1}{{\@currentlabel}{\thepage}}}}}\@gtempa
 \if@nobreak
\ifvmode\nobreak\fi\fi\fi\@esphack}
\gdef\@eqnlabel{#1}}
\def\@eqnlabel{}
\def\@vacuum{}
\def\draftmarginnote#1{\marginpar{\raggedright\scriptsize\tt#1}}
\def\draft{\oddsidemargin
0.0truein
 \def\@oddfoot{\sl preliminary draft \hfil
\rm\thepage\hfil\sl\today\quad\militarytime}
 \let\@evenfoot\@oddfoot
\overfullrule 3pt
 \let\label=\draftlabel
\let\marginnote=\draftmarginnote
\def\@eqnnum{(\theequation)\rlap{\kern\marginparsep\tt\@eqnlabel}
\global\let\@eqnlabel\@vacuum}
}
\catcode`@=12
\def\dj{\hbox{d\kern-0.347em \vrule width 0.3em height 1.252ex depth
-1.21ex \kern 0.051em}}

\def\ee{{\rm e}\,}

\def\pt{\partial}

\def\Dirac{\,\raise.15ex\hbox{/}\mkern-13.5mu D}
\def\dirac{\,\raise.15ex\hbox{/}\kern-.57em \partial}
\def\aslash{\,\raise.15ex\hbox{/}\mkern-13.5mu A}
\def\shalf{{\ifinner {\textstyle \frac{1}{2}}\else \frac{1}{2} \fi}} 
\def\sthreehalfs{{\ifinner {\textstyle \frac{3}{2}}\else \frac{3}{2} \fi}} 
\def\sshalf{{\ifinner {\scriptstyle \frac{1}{2}}\else \frac{1}{2} \fi}} 
\def\sfourth{{\ifinner {\textstyle \frac{1}{4}}\else frac{1}{4} \fi}}
\def\sphifour{{\ifinner {\textstyle \frac{1}{4!}}\else \frac{1}{4!} \fi}}

 \def\CP{{\cal P}}

\def\XXint#1#2#3{{\setbox0=\hbox{$#1{#2#3}{\int}$}
     \vcenter{\hbox{$#2#3$}}\kern-.5\wd0}}

\def\bea{\begin{eqnarray}} \def\eea{\end{eqnarray}}
\def\be{\begin{eqnarray}} \def\ee{\end{eqnarray}} 
\newcommand{\promille}{%
  \relax\ifmmode\promillezeichen
        \else\leavevmode\(\mathsurround=0pt\promillezeichen\)\fi}
\newcommand{\promillezeichen}{%
  \kern-.05em%
  \raise.5ex\hbox{\the\scriptfont0 0}%
  \kern-.15em/\kern-.15em%
  \lower.25ex\hbox{\the\scriptfont0 00}}
\begin{document}

\thispagestyle{empty}

\begin{center}
\hfill IFT-UAM/CSIC-09-10 \\
\hfill UAB-FT-665

\begin{center}

\vspace{1.7cm}

{\LARGE\bf On the Naturalness of Higgs Inflation}
\end{center}

\vspace{1.4cm}

{\bf J.L.F. Barb\'on$^{\,a}$} and {\bf J.R. Espinosa$^{\,b}$}\\

\vspace{1.2cm}

${}^a\!\!$
{\em { Instituto de F\'{\i}sica Te\'orica, IFT-UAM/CSIC, F. Ciencias UAM, 28049 Madrid, Spain}}\\

${}^b\!\!$
{\em {ICREA, Instituci\`o Catalana de Recerca i Estudis Avan\c{c}ats,
Barcelona, Spain}}\\
at {\em { IFAE, Universitat Aut{\`o}noma de Barcelona,
08193 Bellaterra, Barcelona, Spain}}
\\

\end{center}

\vspace{0.8cm}

\centerline{\bf Abstract}
\vspace{2 mm}
\begin{quote}\small
We critically examine the recent claim that the Standard Model Higgs boson
${\cal H}$ could drive inflation in agreement with observations if $|{\cal
H}|^2$ has a strong coupling $\xi\sim 10^4$ to the Ricci curvature scalar.
We first show that the effective theory approach upon which that claim is
based ceases to be valid beyond a cutoff scale $\Lambda=m_p/\xi$, where
$m_p$ is the reduced Planck mass. We then argue that knowing the Higgs
potential profile for the field values relevant for inflation ($|{\cal
H}|>m_p/\sqrt{\xi}\gg \Lambda$) requires knowledge of the ultraviolet
completion of the SM beyond $\Lambda$. In absence of such microscopic
theory, the extrapolation of the pure SM potential beyond $\Lambda$ is
unwarranted and the scenario is akin to other ad-hoc inflaton potentials
afflicted with significant fine-tuning. The appealing naturalness of this
minimal proposal is therefore lost.
\end{quote}

\vfill

\newpage

\section{The Standard Model Higgs as the Inflaton}

\noindent
Following \cite{shap} let us consider the SM Higgs sector corrected by a 
curvature-induced mass term for the Higgs boson:
\be
\label{jordanframe}
\frac{{\cal L_{\rm Jordan}}}{ \sqrt{-g}} = \shalf m_p^2 \,R +\shalf
\xi  h^2  \,R - \shalf  g^{\mu\nu} \,\pt_\mu h \,\pt_\nu h -
\sphifour \lambda (h^2 -v^2)^2\;,
\ee
where $m_p=2.44\times 10^{18}$ GeV is the reduced Planck mass. Here
$h$ stands for the real neutral component of the Higgs doublet ${\cal H}$ that
remains after the  Higgs mechanism, the other scalar degrees of freedom in 
${\cal H}$ 
being absorbed into the longitudinal components of the (omitted)
gauge fields. This parametrization is natural at low energies, and much
less so at energies  well above the symmetry breaking scale, $v=246.22$ GeV.
Nevertheless, discussions  of inflation often incorporate the requirement
of a single-component inflaton field for phenomenological reasons.
Therefore, we will  assume that focusing on a single real component of the 
Higgs field
is appropriate for our discussion at all energies of interest.
 
The effective Lagrangian (\ref{jordanframe}) is assumed to be valid from
the SM scale $v$ up to some threshold below the reduced Planck mass $m_p$.
While the curvature coupling is  irrelevant at low energies, it has quite
remarkable consequences  at large values of the Higgs field \cite{oldies}. 
A suitable
Weyl transformation of the metric
\be
\label{weylresc}
g_{\mu\nu} \rightarrow (1+ \xi h^2 /m_p^2)^{-1}\;g_{\mu\nu}\ ,
\ee
decouples the Higgs field from the curvature operator and  
reveals a new potential 
\be
\label{newpot}
U(h)=\frac{\lambda}{ 4!} \frac{(h^2 - v^2)^2 }{ (1+ \xi h^2 /m_p^2)^2}\;,
\ee
with a modified large-field regime, capable of sustaining slow-roll
inflation. In particular, one finds a plateau for
$h \gg m_p /\sqrt{\xi}$, with energy density
$m_p^4 /\xi^2$. The phenomenological inflationary constraints (small
slow-roll parameters and the right amplitude of density perturbations) can be 
met simply by choosing $\xi \sim 10^4$ \cite{shap} (see also 
\cite{staro,wilczek,develop} for further studies along these lines and 
\cite{effinfl} for some general effective-theory approaches to inflation).

These results suggest that the basic energy scale controlling inflation
is $ \Lambda_I\equiv m_p /\sqrt{\xi}$.   Since $\xi$ is large,  we are
guaranteed that Planck-suppressed effective operators have negligible
effect on the effective potential. 
 In any case, the emergence of the
plateau at large field strength is a remarkable feature that is not at
all expected at the level of the innocent-looking Lagrangian in the
so-called Jordan frame (\ref{jordanframe}). While the occurrence of 
inflationary
plateaus is often attributed to peculiar dynamical features or plain
fine-tuning, we seem to be producing one here, just  from a quite generic 
low-energy effective Lagrangian.

In this note we make two simple observations. First, we show in section 2
that the actual effective cutoff of (\ref{jordanframe}) is $\Lambda = m_p 
/ \xi$, i.e. much
lower than the energy scales determining the properties of the plateau itself.
This suggests that, despite appearances,  there is a strong fine-tuning
implicit in (\ref{newpot}). In connection with this, our second observation
(section 3) is that the plateau is extremely sensitive to the presence of 
operators 
suppressed by the scale $\Lambda$ which are generically expected to appear. 
The claimed naturalness of the scenario is therefore lost.

\section{Effective Cutoff Scale}

\noindent
The presence of nonrenormalizable operators in the effective Lagrangian 
(\ref{jordanframe}) gives an upper bound on its own cutoff scale. 
In the weak-field regime, the operator $h^2 R$ has dimension five by power
counting. Expanding around Minkowski space with a canonically normalized
graviton $g_{\mu\nu} = \eta_{\mu\nu} + m_p^{-1} \gamma_{\mu\nu}$ we find
a leading term\footnote{In a non-zero Higgs background the normalization
of the graviton involves the modified Planck mass $m_p^2+\xi \langle 
h\rangle^2$. This does
not affect qualitatively our results. We also assume $\xi>0$, since  
$\xi<0$  lowers the effective Planck mass
 at large $\langle h\rangle$,  making gravity effects even stronger.}
\be
\label{opel}
\frac{\xi}{m_p} h^2 \eta^{\mu\nu} \pt^2 \gamma_{\mu\nu} + \dots
\ee
where the dots stand for the other tensor structures conforming the
linearized approximation of the Ricci scalar. The contribution of
(\ref{opel}) to tree processes is controlled by the ratio $\xi E/ m_p$ at 
typical
energy $E$,  becoming strongly coupled at $E\sim \Lambda = m_p /\xi$.
Hence the effective cutoff of the nonrenormalizable Lagrangian 
(\ref{jordanframe}) 
is $\Lambda = m_p /\xi$. (Other higher-order operators containing two or 
more gravitons  are suppressed instead by 
$\Lambda_I=m_p/\sqrt{\xi}$ which is a much higher energy scale for $\xi\gg 1$.)

The same effective cutoff  can be obtained in the Einstein frame, in 
which all the non-linearities are transferred to the scalar sector. Both 
the Einstein-frame potential (\ref{newpot})  and the rescaled kinetic 
term of the Higgs field only contain higher-dimensional operators 
suppressed by the naive inflation scale $m_p /\sqrt{\xi}$. There is, 
however, an extra contribution to the two-derivative effective action 
coming from the Weyl rescaling of the curvature, i.e. 
\be
-\frac{3\xi^2}{m_p^2} \frac{h^2}{(1+ \xi h^2 /m_p^2)^2} \,(\pt h)^2
\ee
whose leading term is the dimension-six operator
\be
-3\frac{\xi^2}{m_p^2} \,h^2 \,(\pt h)^2\;,
\ee
with effective cutoff $\Lambda = m_p /\xi$. 

All these nonlinearities of the Higgs kinetic term can be hidden by the field 
redefinition 
\be
\label{redef}
d\phi = d h \,\frac{\left[1+6\xi^2 (h/m_p)^2 + \xi 
(h/m_p)^2\right]^{1/2}}{ 1+ \xi (h/m_p)^2}\;,
\ee
leading to a canonically normalized scalar sector in the Einstein frame,
\be
\label{einsteinframe}
\frac{{\cal L}_{\rm Einstein} }{ \sqrt{-g}} = \shalf m_p^2 \,R -
\shalf (\pt \phi)^2  - U(\phi) \;,
\ee
where the new potential $U(\phi)$ is given by (\ref{newpot}), after we
solve the field redefinition (\ref{redef})  in favor
of the canonical Higgs field  $\phi$ in the Einstein frame.  All couplings appearing in $U(\phi)$ are
non-redundant, since we have exhausted the freedom in nonlinear field redefinitions. 
Working  at small values of the Higgs field we find 
\be
\label{aproxsol}
h = \phi \left[1 - (\xi \phi /m_p)^2 \right] +
{\rm higher \;\;order\;\;terms.}
\ee
Substituting (\ref{aproxsol}) into a generic $h^4$ term of the potential,  
the factor of
$1- \xi^2 \phi^2 / m_p^2 $ induces  a dimension-six effective operator
proportional to 
\be
\frac{\xi^2 }{ m_p^2} \,\phi^6
\;,
\ee
again showing an effective cutoff $\Lambda = m_p /\xi$ (we neglect 
${\cal O}(1)$ factors like $\sqrt{\lambda}$). The same cutoff appears 
after substituting (9) in Higgs Yukawa couplings like that of the top 
quark\footnote{
To clarify further the point already made in our published 
paper let us stress that the operator (10) is the lowest one (in the 
scalar sector) revealing the true cutoff, after all ambiguities due to 
redundant couplings are removed. In this respect, a scattering 
process like $\phi\phi\to\phi\phi\phi\phi$ is more appropriate to discuss 
the breakdown of the effective theory than $\phi\phi\to\phi\phi$. Other 
dimension-6 operators playing a similar role are 
$h_f\phi^3\bar{\psi}_f\psi_f/\Lambda^2$ or 
$g^2\phi^4W_\mu^+W^{\mu -}/\Lambda^2$. As a general comment, for the 
cutoff to be $\Lambda=M_p/\xi$, it is crucial that the Higgs scalar has 
couplings to other sectors or to itself (or the Goldstone components 
of the Higgs doublet). If all those couplings were zero (and the 
Goldstones were absent) the theory in the Einstein frame would correspond 
to a free scalar minimally coupled to gravity (and in the Jordan frame 
that would cause miraculous cancellations in the scattering amplitudes of 
$h$). We thank G. Giudice and A. Riotto for a discussion on 
this point.}. It is  thus 
reassuring that the same high energy cutoff appears
in both frames, as corresponds to a meaningful physical scale 
\cite{catena}. The discussion of 
effective operators in the Einstein frame has the advantage of decoupling 
the problem from
the possible subtleties related to the diffeomorphism invariance in the
gravity sector. 
 
We have confirmed that the natural cutoff of the effective Lagrangian
expanded around the SM region $\phi \sim v$ is given by $\Lambda$ above.
In fact, this piece of limiting information is nearly all we know
about the effective action extrapolated to higher energy scales. Can we trust 
the effective potential for the field values $h>\Lambda_I=m_p/\sqrt{\xi}$ relevant 
for inflation? 
One could argue that quantum fluctuations around that large Higgs background
are very weakly coupled and we do not see there any strong coupling 
effect. However, 
the very existence of the plateau, at energy density 
$m_p^4/\xi^2=\Lambda_I^4\gg\Lambda^4$ is suspect, 
as we discuss next.

\section{UV Sensitivity of the Inflationary Plateau}
\noindent
The field redefinitions carried out so far convert the simple Lagrangian
(\ref{jordanframe}) into a non-polynomial scalar model in terms of the
canonical field $\phi$. From the point of view of effective field theory 
this non-polynomial potential, $U(\phi)$,
is quite fine-tuned though, since all its terms depend  upon a {\it single}
dimensionless parameter $\xi$, apart from the overall constant $\lambda$.
It is then interesting to study the sensitivity of the plateau at
$\phi > m_p /\sqrt{\xi}$ to the presence of higher-dimensional operators
with the same basic structure
of the Jordan frame Lagrangian (\ref{jordanframe}) in the gravitational
sector.

Higher order corrections to the scalar potential in the Jordan frame have
the form
\be
\label{potnop}
V(h) = \sum_{n\geq 4} \frac{1 }{ \Lambda^{2n-4} } \,\lambda_n (h)\,h^{2n}
\;,
\ee
where the effective couplings $\lambda_n$ may depend on  $\log(h/\Lambda)$,
a mild field-dependence induced by radiative corrections to the bare
potential.\footnote{We assume here a mass-independent renormalization scheme
in the effective field theory calculations.}  In general, in the presence
of the coupling (\ref{opel}) all terms in the expansion
(\ref{potnop}) will be generated by radiative corrections, even if
they were not present in the Wilsonian effective potential at the cutoff
scale $\Lambda$. Hence, the low-energy expansion in the vicinity of SM
field strengths suggests that the generic potential has the form\footnote{The
generic mass for $h$ is ${\cal O}(\Lambda)$, a manifestation of the hierarchy
problem, that we do not address here.}
\be
\label{genpott}
V(h) = \Lambda^4 \,{\cal V}(h/\Lambda)\;,
\ee
with ${\cal V}(x)$ an even function of ${\cal O}(1)$, with a typical 
scale of variation
of ${\cal O}(1)$.

Analogous considerations for the curvature term lead to a tower of operators
in the linearized gravity approximation,
\be
\label{towercurv}
\sum_{n\geq 2} \frac{1 }{ \Lambda^{2n-1}} \,\alpha_n (h)\, h^{2n}
\,\eta^{\mu\nu}\pt^2 \gamma_{\mu\nu} + \dots \;,
\ee
where $\alpha_n (h)$ will also have logarithmic field dependence, matched
at the threshold scale $\Lambda$ to the high energy theory.
Using $R\sim m_p^{-1}  \pt^2 \gamma$, we conclude that the generic curvature
coupling at leading order in the Ricci scalar  has the form
\be
\label{gencurvc}
\frac{1}{2}
m_p \,\Lambda \, f(h/\Lambda) \,R\;,
\ee
where, again $f(x)$ is an even function of ${\cal O}(1)$, with typical scale of
variation of ${\cal O}(1)$. While the full effective action below the scale $
\Lambda$ is certainly not exhausted by the functions ${\cal V}(x)$ and $f(x)$, our qualitative points can be 
put forward  by concentrating 
 on those terms directly involved in the inflationary mechanism under discussion.

The Weyl transformation decoupling the curvature from the scalar field is now
\be
\label{weidec}
g_{\mu\nu} \rightarrow \left[1+\frac{\Lambda }{ m_p} f(h/\Lambda)\right]^{-1}
\;g_{\mu\nu}\;,
\ee
and the scalar field redefinition that achieves canonical normalization in
the Einstein frame:
\be
\label{genred}
d\phi = dh \,\frac{\left[1+\xi^{-1} f(h/\Lambda) + 
(3/2)[f'(h/\Lambda)]^2\right]^{1/2}
}{ 1 + \xi^{-1} f(h/\Lambda) }
\;,
\ee
where $f'(x) \equiv df(x)/dx$ and we have defined $\xi \equiv m_p / \Lambda$.
The corresponding scalar potential takes the form
\be
\label{scl}
U(\phi) = \frac{\Lambda^4 \,{\cal V}(h(\phi)/\Lambda) }{ \left[1 + 
\frac{\Lambda }{ m_p}
f(h(\phi)/\Lambda)\right]^2}\;,
\ee
where $h(\phi)$ is the solution of the field redefinition (\ref{genred}).

We can now discuss the general conditions for the emergence of an 
inflationary plateau in the Einstein-frame potential $U(\phi)$. Going 
back to (\ref{genred}), 
there are two
regimes depending on whether $f\ll \xi$ or $f\gg \xi$. In the first case
the dependence on $\xi$ drops out
and we have a generalization of (\ref{aproxsol}), i.e. $h\sim \phi [1+ {\cal O}(\phi^2/\Lambda^2)]$, the weak-field 
regime that
exposes  $\Lambda$ as the 
relevant scale of the problem. In this regime, any plateau must be 
explicitly engineered in the
potential function $V(h)$ from the beginning. 

On the other hand, any unexpected plateau extending over the region 
 $f\gg \xi$ will have the form 
\be
\label{plat}
U(\phi) \approx \Lambda_I^4 \,\frac{{\cal V}(h(\phi)/\Lambda)
}{ [f(h(\phi)/\Lambda)]^2}\;,
\ee
where $\Lambda_I = m_p /\sqrt{\xi}$ is the energy 
scale of the plateau.
Hence, a flat potential at large values of the Higgs field 
requires a functional constraint ${\cal V}(h/\Lambda) \approx 
[f(h/\Lambda)]^2$
over a sufficiently large region so that we sustain enough slow roll. Since the functions ${\cal V}(x)$ and $f(x)$ are
{\it a priori} independent, this constraint is equivalent to the explicit tuning of one function.

 We can be slightly more specific  if we make the extra technical assumption that $f'(x)^2$ is not anomalously small compared to $f(x)$ at large $x$ (which
includes for instance polynomials or trigonometric polynomials). Then we may approximate the field
redefinition (\ref{genred}) as 
\be
\frac{d\phi}{ m_p} \sim \sqrt{\frac{3}{2}} \, \frac{f' (h/\Lambda)
}{ f(h/\Lambda)}\, d(h/\Lambda)\,= \sqrt{\frac{3}{2}} \,d(\log f)\;.
\ee
Hence, we find
\be
f \sim \xi\, \exp\left(a \frac{\phi}{ m_p }\right)
\;\;\;\;{\rm for}\;\;\;f\gg \xi \;,
\ee
where we have defined $a= \sqrt{2/3}$  and we have 
chosen the additive normalization of $\phi$ so that the
matching between the two
regimes, at $f\sim \xi$, corresponds to  $\phi \sim m_p$.

Let us consider the function ${\cal V}(x) / [f(x)]^2$ and eliminate the 
$x$ variable
in favor of the $f$ variable, i.e. we invert $x=x(f)$ and write
\be
\frac{{\cal V}(x) }{ [f(x)]^2} = \frac{{\cal V}(x(f)) }{ f^2}  \equiv F(f)
\;.
\ee
We know that near the origin of field space, and neglecting low-energy
scales of the SM, ${\cal V}\ll \Lambda^4$,
$F(f)$ is a polynomial function with ${\cal O}(1)$ coefficients. In the 
case of
the Lagrangian (\ref{jordanframe}) we have
$ F(f)=1 $ exactly. On the other hand, possible plateaus have
to do with the large $f$ behavior of the function $F(f)$. Any plateau at
large $f$ that ends around $f\sim \xi$ is described by a function that
admits a large $f$ expansion of the form 
\be
F(f) \sim 1+ {\cal O}(\xi /f)\;,
\ee
with $O(1)$ coefficients in general. On the other hand, the complete
potential (\ref{scl}) reads
\be
U= \Lambda_I^4 \frac{F(f) }{ (1+ \xi /f)^2} = \Lambda_I^4 [1+ {\cal 
O}(\xi/f)] =
\Lambda_I^4 \left[1+ {\cal O}\left(e^{-a\phi/m_p }\right)\right]\ ,
\ee 
in the plateau regime. Hence, we have seen that the general picture for 
Higgs inflation 
will come to fruition whenever we have a function $F(f)$ which
approaches unity with power-like accuracy in the region $f\gg  \xi$.
The phenomenological conditions for inflation will be satisfied
just as in ref.~\cite{shap}, with ${\cal O}(1)$ coefficients in all the 
power expansions.
One simply needs to adjust the single large parameter $\xi \sim 10^4$.
So, where is the fine tuning?

The fine-tuning is of course in the assumption that the function $F(f)$
approaches unity at large $f$. Notice that, in general, $F(f)$ cannot even
be defined, because unless $f(x)$ is monotonic, we will not
be able to solve $x$ as a function of $f$ in the first place. For instance,
if ${\cal V}(x)$ and $f(x)$ are `landscape-like' with
a succesion of maxima and minima separated by a distance ${\cal O}(1)$ in 
$x$, we will
not have a plateau in terms of the canonical field. This is very clear in the
particular example  in which we take $f(x)=x^2$
as in ref.~\cite{shap}, and ${\cal V}(x) = \CP(x^2)$ is a pseudoperiodic 
function with
approximate period of ${\cal O}(1)$. The asymptotic form of the potential 
is then
\be
U\rightarrow \Lambda^4 e^{-2a\phi/m_p } \; \CP \left(\xi
e^{a \phi/m_p } \right)\;,
\ee
We see that, in $\phi$ space, we have a singular function with ever
increasing oscillation rate, which in addition is quenched by a decaying
exponential, hardly a good environment for inflation!

From a physical point of view, all we can expect to know is the small field
expansion of the various functions involved, ${\cal V}(x)$ and $f(x)$, 
collected from
fits to amplitudes measured at energies $E\ll \Lambda$. From these 
functions
we obtain $F(f)$, which in this
regime looks like a polynomial in $f$. Suppose we have determined that
it has the form 
\be
\label{expp}
F(f) =  1 + \sum_{n\geq 1} \alpha_n \,f^n\;,
\ee
from fits to low-energy data (fixing the coefficient of $n$-th order
requires measuring effective operators
of dimension $2n$ in the Higgs sector). 
The so-called Higgs inflation will take place if this function approaches
unity with power-like accuracy
as $f> \xi$.  Of course, since $\xi \gg 1$, there is no way we can infer
the behavior at $f/\xi >1$ from
the above low-energy power expansion. If we make an arbitrary truncation 
at $n\leq N$
(this will always be the case because the experimental accuracy will be finite) and
extrapolate the result to large $f$, we either need miraculous
cancellations, or else we must bound
\be
\label{ftune}
|\alpha_n | \ll \frac{1}{ \xi^n}\;,
\ee
for all $n\leq N$, which is a pretty strong fine-tuning indeed. In other 
words, if one starts 
with a generic potential without plateau in the Einstein frame, the 
potential transformed back 
into the Jordan frame through eqs.~(\ref{weylresc}) and (\ref{redef}) 
will look at low-energy 
like a normal SM Higgs potential plus a tower of non-renormalizable 
operators difficult to measure.
Without detailed information about the structure of that tower of 
operators one cannot predict the
presence of a plateau in the Einstein frame.

It is interesting to reinterpret this fine-tuning in terms of scales again.
Let us set $f(x) = x^2$, so that
(\ref{expp}) translates into an expansion of the dimensionless potential 
function:
\be
{\cal V} = f^2 + \sum_{n\geq 1} \alpha_n \,f^{n+2} = x^4 + \sum_{n\geq 1}
\alpha_n \,x^{2n+4}\;,
\ee
which in turn translates into the following expansion of the  Jordan-frame
potential
\be
V(h) = h^4 + \sum_{n\geq 1} \alpha_n \frac{h^{2n+4} }{ \Lambda^{2n}} = 
h^4 + \sum_{n\geq 1} \beta_n \frac{h^{2n+4}} { \Lambda_I^{\,2n}}
\;,
\ee
where we have written $\beta_n =\alpha_n\ \xi^n$. The very stringent 
fine-tuning condition (\ref{ftune})  actually means that the potential is 
`natural' with respect to the higher scale $\Lambda_I= 
\sqrt{\xi}\,\Lambda$, with at most a slow-roll fine-tuning \footnote{By 
slow-roll fine-tuning we mean the `technical' tuning needed to keep the 
slow-roll parameters small.}  of the coefficients 
$|\beta_n | \ll 1$.

Therefore,  in this example Higgs inflation is supported by the fact 
that, while the non-minimal curvature coupling in the Jordan frame is 
heralding the presence of an effective cutoff at the scale $\Lambda$, the 
effective scalar potential in the same frame has a hierarchically larger 
effective cutoff  $\Lambda_I$.  However, even this situation is not enough 
to guarantee the existence of the plateau, since any significant 
deviation from the behavior $f(x)=x^2$ at large values of  $f\sim \xi$ 
will destroy the plateau as well. 

We conclude that a very peculiar structure of the effective action is 
required, with a hierarchical separation of scales within different 
sectors, combined with a {\it strong correlation} of these sectors at 
field strengths
in excess of the  {\it higher} scale. Even if one such effective action 
is set at the cutoff scale $\Lambda$, radiative corrections from quantum 
loops below $\Lambda$ will generically upset such fragile structure. 
Needless to say, this problem afflicts also other inflationary models
that use the same large coupling of the inflaton to the Ricci scalar as a 
way of obtaining a plateau
without invoking any symmetry reason.

\section*{Acknowledgments}

\noindent
We thank Guillermo Ballesteros, Gian Giudice and Antonio Riotto for
discussions. This work was partially supported by the European
Commission  through the Marie Curie Research
and Training Networks ``Quest for Unification" (MRTN-CT-2004-503369), 
``ForcesUniverse"  (MRTN-CT-2004-005104) 
and ``UniverseNet" (MRTN-CT-2006-035863); by the Spanish
Consolider-Ingenio 2010 Programme CPAN (CSD2007-00042);   the Comunidad
Aut\'onoma de Madrid under grant HEPHACOS P-ESP-00346 and the Spanish 
Ministry MICNN under contracts  
FPA2006-05485 and FPA 2007-60252. 

\section*{Note added}

While this paper was being drafted, ref.~\cite{burgess} appeared. In this work
related issues are discussed, including the identification of a low
effective cutoff scale $m_p /\xi$, and the possible pernicious influence
of higher non-renormalizable Higgs operators (coming from integrating out
some heavy sector coupled to the Higgs) on the potential plateau.  While
the authors of ref.~\cite{burgess} examine the applicability of the
semiclassical approximation and leave open the door for a (narrow)
parameter region where inflation could be viable, we emphasize here that
the very existence of the inflationary plateau is compromised by the
result $\Lambda = m_p /\xi$.


\begin{thebibliography}{10}
%
%
%\cite{Bezrukov:2007ep}
\bibitem{shap}
  F.~L.~Bezrukov and M.~Shaposhnikov,
  %``The Standard Model Higgs boson as the inflaton,''
  Phys.\ Lett.\  B {\bf 659}, 703 (2008)
  [hep-th/0710.3755];
  %%CITATION = PHLTA,B659,703;%%
%\cite{Bezrukov:2008ej}
%\bibitem{Bezrukov:2008ej}
  F.~L.~Bezrukov, A.~Magnin and M.~Shaposhnikov,
  %``Standard Model Higgs boson mass from inflation,''
  [hep-ph/0812.4950];
  %%CITATION = ARXIV:0812.4950;%%
%\cite{Bezrukov:2008ut}
%\bibitem{Bezrukov:2008ut}
  F.~Bezrukov, D.~Gorbunov and M.~Shaposhnikov,
  %``On initial conditions for the Hot Big Bang,''
  [hep-ph/0812.3622].
  %%CITATION = ARXIV:0812.3622;%%
%
%
%\cite{Salopek:1988qh}
\bibitem{oldies}
  D.~S.~Salopek, J.~R.~Bond and J.~M.~Bardeen,
  %``Designing Density Fluctuation Spectra in Inflation,''
  Phys.\ Rev.\  D {\bf 40}, 1753 (1989);
  %%CITATION = PHRVA,D40,1753;%%
%\cite{Fakir:1990eg}
%\bibitem{Fakir:1990eg}
  R.~Fakir and W.~G.~Unruh,
  %``Improvement on cosmological chaotic inflation through nonminimal
  %coupling,''
  Phys.\ Rev.\  D {\bf 41}, 1783 (1990);
  %%CITATION = PHRVA,D41,1783;%%
%
%\cite{Kaiser:1994vs}
%\bibitem{Kaiser:1994vs}
  D.~I.~Kaiser,
  %``Primordial spectral indices from generalized Einstein theories,''
  Phys.\ Rev.\  D {\bf 52}, 4295 (1995)
  [astro-ph/9408044];
%  
%%CITATION = PHRVA,D52,4295;%%
%\cite{Komatsu:1999mt}
%\bibitem{Komatsu:1999mt}
  E.~Komatsu and T.~Futamase,
  %``Complete constraints on a nonminimally coupled chaotic inflationary
  %scenario from the cosmic microwave background,''
  Phys.\ Rev.\  D {\bf 59}, 064029 (1999)
  [astro-ph/9901127].
  %%CITATION = PHRVA,D59,064029;%%
%
%
%\cite{Barvinsky:2008ia}
\bibitem{staro}
  A.~O.~Barvinsky, A.~Y.~Kamenshchik and A.~A.~Starobinsky,
  %``Inflation scenario via the Standard Model Higgs boson and LHC,''
  JCAP {\bf 0811}, 021 (2008)
  [hep-ph/0809.2104].
  %%CITATION = JCAPA,0811,021;%%
%
%
%\cite{DeSimone:2008ei}
\bibitem{wilczek}
  A.~De Simone, M.~P.~Hertzberg and F.~Wilczek,
  %``Running Inflation in the Standard Model,''
  [hep-ph/0812.4946].
  %%CITATION = ARXIV:0812.4946;%%
%
%
\bibitem{develop}
%\cite{Park:2008hz}
%\bibitem{Park:2008hz}
  S.~C.~Park and S.~Yamaguchi,
  %``Inflation by non-minimal coupling,''
  JCAP {\bf 0808} (2008) 009
  [hep-ph/0801.1722];
  %%CITATION = JCAPA,0808,009;%%
%\cite{Bauer:2008zj}
%\bibitem{Bauer:2008zj}
  F.~Bauer and D.~A.~Demir,
  %``Inflation with Non-Minimal Coupling: Metric vs. Palatini Formulations,''
  Phys.\ Lett.\  B {\bf 665} (2008) 222
  [hep-ph/0803.2664];
  %%CITATION = PHLTA,B665,222;%%
%\cite{Kaloper:2008gs}
%\bibitem{Kaloper:2008gs}
  N.~Kaloper, L.~Sorbo and J.~Yokoyama,
  %``Higgsflation at the GUT scale in a Higgsless Universe,''
  Phys.\ Rev.\  D {\bf 78} (2008) 043527
  [hep-ph/0803.3809];
  %%CITATION = PHRVA,D78,043527;%%
%\cite{Hrycyna:2008gk}
%\bibitem{Hrycyna:2008gk}
  O.~Hrycyna and M.~Szydlowski,
  %``Dynamics of extended quintessence on the phase plane,''
  [hep-th/0812.5096];
  %%CITATION = ARXIV:0812.5096;%%
%\cite{GarciaBellido:2008ab}
%\bibitem{GarciaBellido:2008ab}
  J.~Garcia-Bellido, D.~G.~Figueroa and J.~Rubio,
  %``Preheating in the Standard Model with the Higgs-Inflaton coupled to
  %gravity,''
  [hep-ph/0812.4624].
  %%CITATION = ARXIV:0812.4624;%%
%
%
%\cite{Burgess:2003jk}
\bibitem{effinfl}
  C.~P.~Burgess,
  %``Quantum gravity in everyday life: General relativity as an effective field
  %theory,''
  Living Rev.\ Rel.\  {\bf 7} (2004) 5
  [gr-qc/0311082];
  %%CITATION = 00222,7,5;%%
%\cite{Boyanovsky:2005pw}
%\bibitem{Boyanovsky:2005pw}
  D.~Boyanovsky, H.~J.~de Vega and N.~G.~Sanchez,
  %``Clarifying inflation models: Slow-roll as an expansion in 1/N(efolds) and
  %no fine tuning,''
  Phys.\ Rev.\  D {\bf 73} (2006) 023008
  [astro-ph/0507595];
  %%CITATION = PHRVA,D73,023008;%%
%\cite{Boyanovsky:2009xh}
%\bibitem{Boyanovsky:2009xh}
  D.~Boyanovsky, C.~Destri, H.~J.~de Vega and N.~G.~Sanchez,
  %``The Effective Theory of Inflation in the Standard Model of the Universe and
  %the CMB+LSS data analysis,''
  [astro-ph/0901.0549].
  %%CITATION = ARXIV:0901.0549;%%
%
%
%\cite{Catena:2006bd}
\bibitem{catena}
  R.~Catena, M.~Pietroni and L.~Scarabello,
  %``Einstein and Jordan frames reconciled: a frame-invariant approach to
  %scalar-tensor cosmology,''
  Phys.\ Rev.\  D {\bf 76} (2007) 084039
  [astro-ph/0604492].
  %%CITATION = PHRVA,D76,084039;%%
%
%
%\cite{Burgess:2009ea}
\bibitem{burgess}
  C.~P.~Burgess, H.~M.~Lee and M.~Trott,
  %``Power-counting and the Validity of the Classical Approximation During
  %Inflation,''
  [hep-ph/0902.4465].
  %%CITATION = ARXIV:0902.4465;%%
%
%
\end{thebibliography}
\end{document}